\documentclass{WileyMSP-template}
\usepackage{soul}
\usepackage{color}
\usepackage[utf8]{inputenc}
\usepackage{setspace}
\usepackage{subcaption}
\usepackage{amsmath}
\usepackage{dsfont}
\usepackage{lineno}
\usepackage[dvipsnames]{xcolor}
\usepackage[normalem]{ulem}

\usepackage{chngcntr}

\usepackage{graphicx} 

\begin{document}

\pagestyle{fancy}
\rhead{\includegraphics[width=2.5cm]{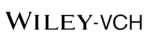}}

\title{Neuromorphic Photonic Processing and Memory with Spiking Resonant Tunnelling Diode Neurons and Neural Networks}

\maketitle


\author{Dafydd Owen-Newns}
\author{Joshua Robertson}
\author{Giovanni Donati}
\author{Jos\'{e} Figueiredo}
\author{Edward Wasige}
\author{Kathy L\"{u}dge}
\author{Bruno Romeira}
\author{Antonio Hurtado*}

\begin{affiliations}
A. Hurtado\\
Institute of Photonics, University of Strathclyde, Glasgow, United Kingdom\\
Email Address: antonio.hurtado@strath.ac.uk

\end{affiliations}


\keywords{Neuromorphic photonics, Resonant Tunnelling Diodes, Photonic Neurons, Photonic Spiking Neural Networks}

\begin{abstract}

Neuromorphic computing—modelled after the functionality and efficiency of biological neural systems—offers promising new directions for advancing artificial intelligence and computational models. Photonic techniques for neuromorphic computing hardware are attracting increasing research interest, thanks to their potentials for ultra high bandwidths, low-crosstalk and high parallelism. Among these, approaches based upon resonant tunnelling diodes (RTDs) have recently gained attention as potential building blocks for next-generation light-enabled neuromorphic hardware, due to their capacity to replicate key neuronal behaviours such as excitable spiking and refractoriness, added to their potentials for high operational speeds, energy efficiency and compact footprints. In particular, their ability to function as opto-electronic spiking neurons makes them strong candidates for integration into novel event-based neuromorphic computing systems. This work demonstrates the application of optically-triggered spiking RTD neurons to a multiplicity of applications and architectures, these include systems based upon single elements for multi-modal (photonic-electronic) fast rising edge-detection in time-series data, the construction of a two-layer feedforward artificial photonic spiking neural network (pSNN) using RTD neurons as the nonlinear nodes delivering excellent performance in complex dataset classification tasks, and a pSNN comprised of multiple coupled light-sensitive RTD spiking neurons that supports performance as an adjustable neuromorphic optical spiking memory system with a tunable storage time of spiking patterns.

\end{abstract}


\section{Introduction}

As the volume of global data continues to grow exponentially, machine learning (ML) and artificial intelligence (AI) have rapidly advanced, becoming pervasive across virtually every sector. These technologies have enabled machines to perform complex cognitive tasks such as learning, computer vision, natural language processing, and sophisticated pattern recognition—tasks once thought to be uniquely human. However, while software algorithms have seen dramatic breakthroughs, the hardware that supports them often receives far less attention. Today’s large-scale ML models are typically trained on energy-intensive, cloud-based computing clusters. Remarkably, the energy required to train a single state-of-the-art model has been estimated to rival the total energy consumption of a human brain over six years. This rising computational demand, coupled with diminishing returns from traditional chip scaling, has sparked increasing interest in alternative, more energy-efficient computing paradigms. Among these, neuromorphic engineering stands out. Inspired by the structure and function of the brain, neuromorphic systems aim to replicate the computational efficiency of biological neurons. These architectures vary in their degree of biological realism—from conventional implementations of artificial neural networks (ANNs) on specialized hardware to models that emulate the dynamic behaviour of real neurons more closely. Examples of large scale neuromorphic computers include SpiNNaker and SpiNNaker 2 from the University of Manchester and Technische Universitat Dresden \cite{furber2014spinnaker, höppner2022spinnaker2}, IBM’s TrueNorth and Northpole \cite{debole2019truenorth,Modha2023Northpole}, and Intel’s Loihi and Loihi 2 chips \cite{davies2018loihi,Orchard2021Loihi2}, each of which run on specialised integrated electronic hardware.

Photonic platforms present compelling opportunities for advancing neuromorphic computing by capitalizing on the inherent strengths of optical systems, such as ultra-high bandwidth, minimal signal interference, and the ability to support long-range communication links. These features position photonics as a promising route for overcoming critical bottlenecks faced by conventional electronics based architectures \cite{Brunner2025}. Notably, optical systems support high-throughput data processing, enabling dense and massively parallel computation via mechanisms like wavelength-division and mode-division multiplexing. Moreover, photonics enables access to a diverse set of linear and nonlinear optical effects, which can be exploited efficiently for complex signal processing and computational tasks. The optical domain also holds distinct advantages in terms of energy efficiency and speed, with the potential for low-power interconnects and inherently faster operation rates, making it a strong candidate for next-generation neuromorphic platforms. Recent implementations have demonstrated optical neural network components utilizing devices such as modulators \cite{ZhangH2021}, micro-ring resonators \cite{Giron2024,Donati2022}, phase-change materials \cite{Feldmann2019}, and semiconductor lasers\cite{Zheng2023Edoc}.

Resonant Tunnelling Diodes (RTDs) have recently drawn increasing attention for their ability to act as opto-electronic spiking neurons \cite{Jacob2025, Zhang2024}. RTDs are semiconductor devices that incorporate into their structure a double barrier quantum well that is sufficiently small ($\approx$ 10\,nm wide) to allow the tunnelling of electrons. Due to the double-barrier structure, at particular energies of the incident electrons the tunnelling probability can reach 100\%, a phenomenon known as resonant tunnelling. This gives rise to the highly nonlinear N-shaped current-voltage relation, characteristic of RTDs, which shows a localised current peak (that occurs at resonant tunnelling) followed by a region of negative differential resistance (NDR), as tunnelling moves off-resonance. The speed of electron tunnelling combined with this highly nonlinear I-V curve has found RTDs uses as ultrafast oscillators (operating at THz frequencies) for telecommunications \cite{cimbri2022resonant, nishida2019terahertz} and in recent years in neuromorphic technologies for operation as photonic spiking neurons\cite{robertson2025leaky}, as these can exhibit behaviours analogous to those of biological neurons such as excitable spiking \cite{ZhangW:PRTD,Jacob2025}, spike bursting \cite{PhysRevApplied.15.034017,Donati2024}, and integrate-and-fire \cite{Robertson2024Utip} mechanisms. Previous works have demonstrated the use of RTD neurons for photonic neuromorphics by coupling the RTD directly to photonic elements such as photodetectors (for receiving optical signals) and laser diodes (for transmitting optically). RTDs have been built which contain in their structure a photodetecting layer that can directly convert light incident on the RTD into input current \cite{PhysRevApplied.17.024072,Qusay2023,Jacob2025}.

This work will introduce several new applications of light-sensitive spiking RTD neurons for photon-enabled neuromorphic processing of data based upon different architectures. First, we demonstrate experimentally and in theory that single photonic-electronic RTD neurons can be used to perform multi-modal processing tasks at high-speeds, such as the detection of fast rising edges in time-series data; thus, highlighting the excellent computational capabilities of these devices. Moreover, we also expand our theoretical investigations to neural network architectures of light-sensitive RTD spiking neurons for novel applications with increased complexity. First, we use multiple uncoupled RTD neurons for the construction of a spatially-multiplexed photonic spiking neural network (SNN), based upon the extreme learning machine (ELM) paradigm, demonstrating its successful operation in complex dataset classification tasks at ultrafast speeds. Furthermore, we also report on networks of coupled RTD spiking neurons allowing operation as an adjustable neuromorphic optical memory system. 

\section{Spiking Resonant Tunnelling Diode Neurons}

RTD devices feature a characteristic double-barrier quantum well (DBQW) in their epilayer semiconductor structure that enables resonant electron tunnelling and a highly nonlinear N-shaped current-voltage relationship with a region of negative differential resistance. Notably, when biased inside (or at the boundaries of) this NDR region, RTDs can elicit excitability (spike firing regimes) analogous to those in biological neurons, but multiple orders of magnitude faster. Importantly, RTDs can also incorporate photo-detection layers in their stacked structure, making them sensitive to light, enabling their use as photonic (as well as electronic) spiking neurons \cite{PhysRevApplied.17.024072}. These properties signpost RTDs a very interesting technology platform for multi-modal (photonic-electronic) neuromorphic, spike-based, hardware \cite{Qusay2023}. In this work, we provide experimental and numerical analyses of photonic-electronic RTD-based neuromorphic systems. 

\begin{figure}[t]
    \centering
    \includegraphics[width=0.95\linewidth]{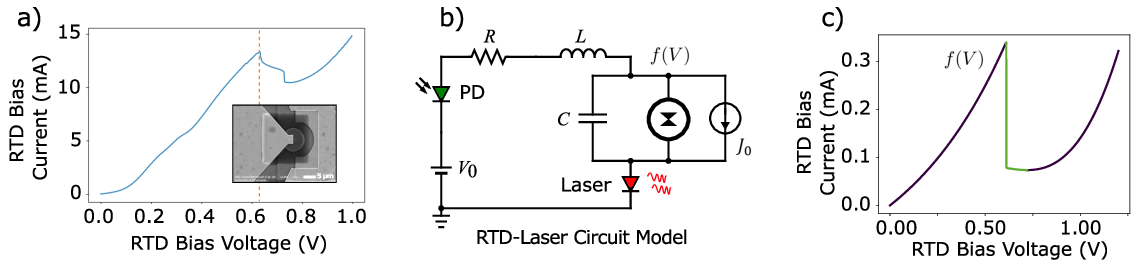}
    \caption{(a) I-V curve of the experimental RTD showing a NDR region between the so-called peak (dotted-line) and valley points (0.62\,V and 0.7\,V, respectively). Inset, an SEM of a fully fabricated photodetecting RTD (5\,$\mu$m scale-bar). (b) Lumped circuit model of the RTD spiking neuron connected to laser (RTD-laser). The voltage source $V_0$ is representative of both radio-frequency (RF) modulation and direct-current (DC) bias. The resistance and inductance parameters are intrinsic parasitic features of all electrical connections in the system, and the equivalent capacitance is a feature of the DBQW in the RTD. (c) The I-V curve of the simulated RTD of this work, with an NDR region from 0.6\,V to 0.72\,V highlighted.}
    \label{fig:circuit}
\end{figure}

Experimentally, we employ a light-sensitive photo-detecting RTD, with a 500\,nm mesa radius nanopillar, and a 9\,$\mu$m diameter optical window. Full details of the experimental RTD device and its heterostructure composition can be found in previous reports \cite{Qusay2023}. The design of the RTD enables its use as a light-sensitive detector at infrared wavelength ranges, including the 1300\,nm and 1550\,nm wavelength regions. \textbf{Figure \ref{fig:circuit}}(a) plots a scanning electron microscope (SEM) image of the experimental photo-detecting RTD device and its measured I-V relationship. Figure \ref{fig:circuit}(a) reveals the existence of a region of negative differential resistance ranging from approx. 0.62\,V to 0.7\,V, between the so-called peak and valley bias voltage points. This device was able to exhibit deterministic excitable spike firing regimes when externally perturbed, both electronically and optically \cite{PhysRevApplied.17.024072,ZhangW:PRTD}, permitting its use as multi-modal artificial spiking neuron and test-bed for the theoretical model.

Numerically, we model the light-enabled Resonant Tunnelling Diode spiking neuron by considering the system as a lumped circuit, as shown in Figure \ref{fig:circuit}(b), and as described in \cite{PhysRevApplied.17.024072}. Using Kirchhoff's circuit laws, the dynamics of the voltage across, and current through, the RTD can be described by \textbf{Equation \ref{eq:Vf}} and \textbf{\ref{eq:If}}, shown below. 

\begin{align}
    \label{eq:Vf}
    C\dot{V} &= I - f(V)-\kappa S_0(t) \\
    \label{eq:If}
    L\dot{I} &= V_0(t) - V - RI
\end{align}

where $V$ and $I$ are the voltage and current respectively, $V_0(t)$ is the supplied voltage (including RF modulation and DC bias), $S_0(t)$ is the intensity of light incident on the photo-detecting layer of the RTD (with current conversion factor $\kappa$), and $R$, $L$, and $C$ are circuit equivalent resistance, capacitance and inductance, respectively. $f(V)$ is the voltage-dependent conductivity of the RTD \cite{RTD_IV}, as described in \textbf{Equation \ref{eq:f_IV}}: 

\begin{equation}
\label{eq:f_IV}
    \begin{aligned}
        f(V) = &A\ln\left[\frac{1+e^{(q/k_BT)(b-c+n_1V)}}{1+e^{(q/k_BT)(b-c-n_1V)}}\right]\\
        &\times\left[\frac{\pi}{2}+tan^{-1}\left(\frac{c-n_1V}{d}\right)\right]+H\left[e^{(q/k_BT)n_2V}-1\right]
    \end{aligned}
\end{equation}

where $A$, $b$, $c$ ,$d$, $n_1$, $n_2$, and $H$ are parameters used to fit and match the features of modelled RTD to that of the experimentally measured RTDs, as described in \cite{PhysRevApplied.17.024072}. \textbf{Table \ref{tbl:params}} collects the selected parameter values used in this work. These parameters were chosen as in \cite{PhysRevApplied.17.024072} to fit an experimentally determined I-V curve

\begin{table}[h]
  \caption{Parameters used in the RTD model}
  \begin{tabular}[htbp]{@{}lll@{}}
    \hline
    Parameter Name & Value\\
    \hline
    $R$  & $10\,\Omega$ \\
    $L$  & $126\,nH$ \\
    $C$ &  $0.002\,pF$ \\
    $\kappa$ & $0.1\times10^{-6}\,A$ \\
    $A$ & $-5.5\times10^{-5}\,A$ \\
    $b$ & $0.033\,V$ \\
    $c$ & $0.113\,V$ \\
    $d$ & $-2.8\times10^{-6}\,V$ \\
    $n_1$ & $0.185$ \\
    $n_2$ & $0.045$ \\
    $H$ & $18\times10^{-5}\,A$\\ 
    \hline
  \end{tabular}
  \label{tbl:params}
\end{table}

Figure \ref{fig:circuit}(c) shows the modelled I-V curve of the RTD described by $f(V)$. The two inversion points at the local maximum (peak) and minimum (valley) of the I-V curve occur at the voltage values of $0.6V$ and $0.72V$. It is precisely around the peak and valley points that the RTD must be biased in order to produce excitable spiking regimes (under both electronic or optical excitation) via transitions in and out of the NDR \cite{PhysRevApplied.15.034017}. The parameters selected during the modelling of the RTD were purposely chosen to highlight the potential for the high-frequency ($>$\,GHz), energy-efficient operation ($\sim$\,100\,pJ/spike) of RTD technology \cite{PhysRevApplied.17.024072}. Therefore, whilst the voltage boundaries of the NDR region achieve similar values to the experimental RTD, the numerical results realise higher operation speeds due to parasitic setup limitations in the experimental work \cite{PhysRevApplied.15.034017,PhysRevApplied.17.024072}.

Additionally, since RTDs can be externally coupled to light sources, such as semiconductor lasers \cite{PhysRevApplied.17.024072}, we convert the excitable spikes from the electrical domain to the optical domain; thus, effectively creating an optical I/O spiking neuron system. Here, in our modelling, we include for completeness a simple two-level laser rate equation model \cite{10.1063/1.5022958} for a semiconductor laser, defined by \textbf{Equation \ref{eq:Nf}} and \textbf{\ref{eq:Sf}} below.

\begin{align}
    \label{eq:Nf}
    \dot{N} &= \frac{J-\eta I}{q}-(\gamma_m+\gamma_l+\gamma_{nr})N - \gamma_m(N-N_0)S \\
    \label{eq:Sf}
    \dot{S} &= \left(\gamma_m(N-N_0)-\frac{1}{\tau_{ph}}\right)S + \gamma_mN
\end{align}

In Equation \ref{eq:Nf} and \ref{eq:Sf}, $N$ and $S$ represent the carrier inversion and photon number. $J$ is the bias current, $\eta$ is the coupling efficiency, $\tau_{ph}$ is the lifetime of photons in the cavity, $N_0$ is the threshold inversion number and the constants $\gamma_m$, $\gamma_l$ , $\gamma_{nr}$ are the spontaneous emission rate, radiative leaking rate and non-radiative leaking rate, respectively. To couple the output of the modelled RTD to the laser model, and create an RTD-laser, the current through the RTD ($I$), is used to directly drive the laser with the coupling efficiency factor $\eta$. The spiking dynamics and excitability properties of this RTD-laser system are described in \cite{ZhangW:PRTD,PhysRevApplied.15.034017,PhysRevApplied.17.024072}, demonstrating the response of the RTD circuit to perturbations of bias voltage $V_0$ and of the intensity of incident light $S_0$, either individually or simultaneously. in this work, Equation 1-5 were solved numerically using a fourth-order implicit Gauss-Legendre \cite{IserlesA1996Afci} integrator.

\section{Results}

\subsection{Neuromorphic Processing with single RTD neurons: Edge-feature detection in time-series }
\label{sec:ed}

In this Section, we investigate the capability of single RTD spiking neurons, showcasing simultaneous operation with multi-modal data inputs (photonic and electronic). Specifically, using Equation \ref{eq:Vf}-\ref{eq:f_IV}, we model an RTD spiking neuron acting as a high-speed processing element, able to detect (at very fast rates) edge-features in complex time-series data. We initially investigate this behaviour numerically, before testing the time-series processing task experimentally with a photo-detecting RTD neuron operating at telecom infrared wavelengths (in the 1550\,nm window). 

\begin{figure}
    \centering
    \includegraphics[width=0.8\linewidth]{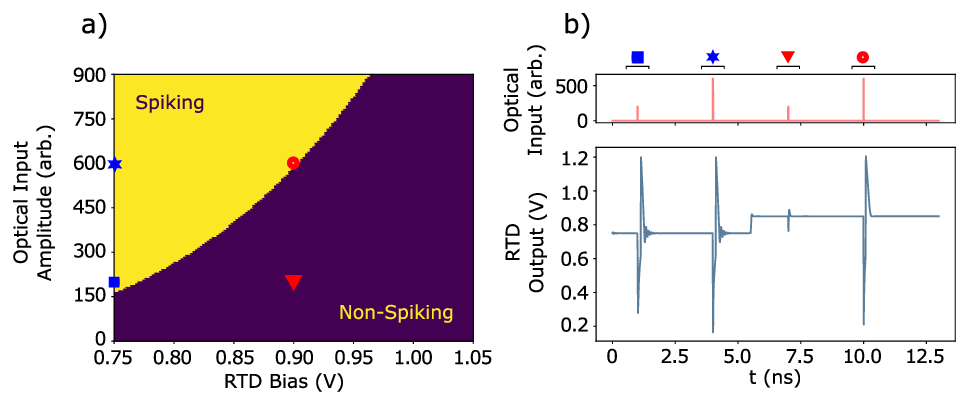}
    \caption{Numerical analysis of multi-modal (photonic-electronic) spiking regimes in an RTD neuron. (a) Optical spike firing threshold amplitude (in arbitrary simulated units) against bias voltage. As the bias on the RTD is increased, the amplitude of an incoming optical pulse must be higher in order to trigger a spike response in the system (yellow pixels represent a spike being triggered, while purple represents no spike). (b) Simulated optical inputs of 20\,ps-long pulses of different intensities injected into an RTD (top, red time traces). Simulated output voltage of an RTD neuron's response (bottom, blue time-series) when the device is based at two voltages, 0.75\,V and 0.85\,V. Respective symbols indicate the location of input conditions on the map.}
    \label{fig:ed:thr}
\end{figure}

An RTD, when biased in the ‘valley’ region of its I-V curve, can be made to fire a spike by an incoming perturbation (i.e. an optical/electrical pulse) with strength above the device’s excitation threshold. This is shown graphically in \textbf{Figure \ref{fig:ed:thr}}. The theoretical parameter map shown in Figure 
 \ref{fig:ed:thr}(a) reveals combinations of applied bias voltages and input optical pulse amplitudes (perturbations) for which the RTD neuron will (or not) fire an excitable spike upon the arrival of a 20ps-long input. Regions of spike firing and quiescence are depicted respectively by yellow and blue regions in the map (for optical pulses of different duration, these regions will be different, particularly the minimum amplitude that will trigger spikes will be lower for higher duration pulses). Figure \ref{fig:ed:thr}(b) plots in turn the numerically simulated output time-series of the RTD neuron when operated with bias voltages of $0.75\,V$ and $0.85\,V$, for different optical input amplitudes. The simulated conditions are indicated on the map using blue and red symbols (see Figure \ref{fig:ed:thr}(a)). Figure \ref{fig:ed:thr}(b) demonstrates that these two system parameters can controllably tune the threshold for spike firing in the RTD neuron, highlighting its computational capability and multi-modal (photonic and electronic) operation. A similar effect occurs when the RTD is biased in its ‘peak’ region, where excitable spikes can also be triggered, subject to input perturbations of reverse polarity (drops in optical optical power) \cite{PhysRevApplied.17.024072}. In RTD neurons, dual (photonic-electronic) modulation therefore makes it possible to combine information from two heterogeneous signals and produce a neuromorphic spiking signal as an output. 

In this work, photonic-electronic dual modulation is used to perform event-based detection of rising edge features in time-series inputs. Here, the RTD is programmed to fire excitable spikes when the amplitude of a raw signal increases by a sufficient amount over a short period of time, i.e. when the difference $D_\tau$ (according to \textbf{Equation \ref{eq:diff}}) exceeds some value.

\begin{equation}
\label{eq:diff}
    D_{\tau}(t) = y(t) - y(t-\tau),
\end{equation}

\begin{figure}
    \centering
    \includegraphics[width=0.65\linewidth]{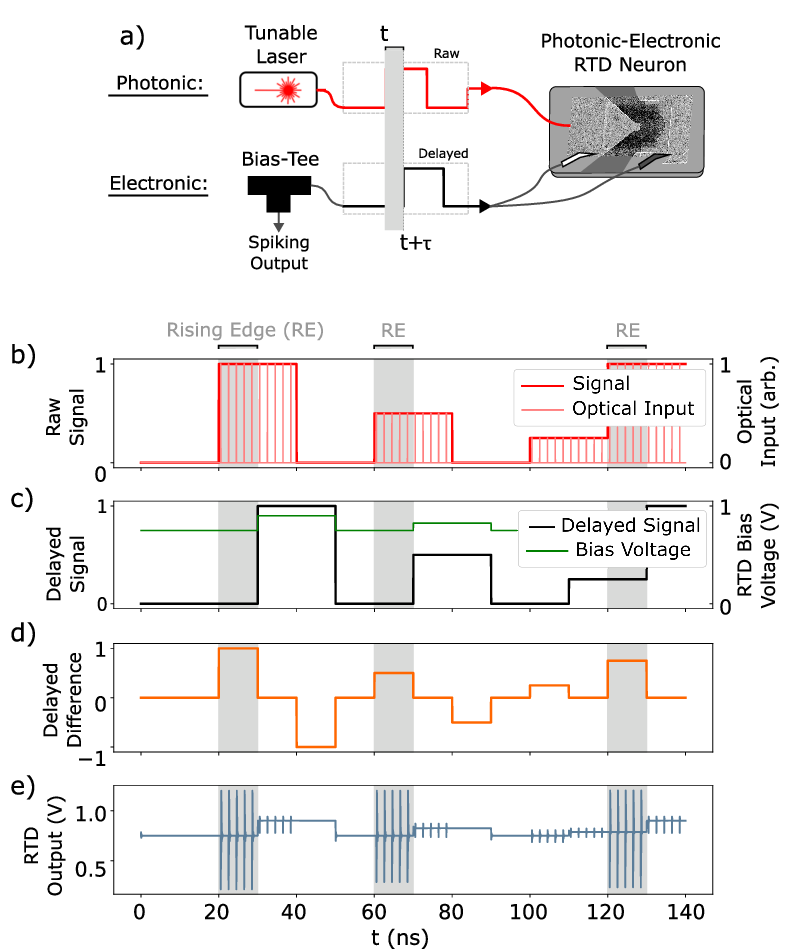}
    \caption{Operation principle of RTD event-based time-series rising edge detection. (a) Schematic of a photonic-electronic RTD neuron subject to simultaneous modulation of its optical input and bias voltage. (b) An example raw signal is simulated for the purpose of the demonstration (red line). The raw signal amplitude is encoded into short (20\,ps) rectangular optical pulses (at 0.5\,GHz) to optically modulate the RTD neuron (pink line). (c) A 10\,ns-delayed copy of the raw time-series signal (black line) is also electrically encoded into the RTD by modulating its voltage bias. An amplitude modulation of up to 0.15\,V is performed on-top of the 0.75\,V constant bias to push the RTD operation point away from the valley region of its I-V curve (green line). (d) The calculated amplitude difference between the raw signal and its 10\,ns-delayed copy (orange line). The delay difference visualises the simultaneous, counteracting, photonic-electronic inputs and reveals a clear threshold above which the RTD neuron will fire. (e) Calculated temporal output from an RTD subject to photonic-electronic modulation (blue line), demonstrating successful edge detection via fast spike-firing events (marked with grey shadow regions).}
    \label{fig:ed:example}
\end{figure}

Raw time-series values $y(t)$ are combined with a delayed signal copy $y(t-\tau)$ to create $D_\tau(t)$), that when high can be setup to exceed the spike-firing threshold of an RTD neuron. To achieve this, we first encoded the raw time-series value ($y(t)$) into the amplitude of a rectangular optical pulse. This optical pulse was then injected into the RTD neuron, modulating the device with a photonic signal. The delayed time-series value ($y(t-\tau)$) was then used to electrically control the bias of the RTD, modulating dynamically the spiking threshold of the RTD neuron. \textbf{Figure \ref{fig:ed:example}}(a) illustrates graphically the operating principle of the RTD event-based time-series edge detection system. 

Figure \ref{fig:ed:example}(b) and \ref{fig:ed:example}(c) plot the simulated input time-series signal and its delayed copy, while Figure \ref{fig:ed:example}(d) plots their respective amplitude difference. The optical input signal is created using 20\,ps-long pulses repeated every 2\,ns. The maximum amplitude of the optical input was 600 arbitrary units, as used in Figure \ref{fig:ed:thr}(b), scaled by the input in Figure \ref{fig:ed:example}(b). The delayed copy is created using a 0.75\,V bias voltage and a modulation voltage of up to 0.15\,V, creating a maximum momentary voltage of 0.9\,V during the peak of the delayed time-series signal (Figure \ref{fig:ed:example}(c)). Figure \ref{fig:ed:example}(e) shows the simulated temporal response from the RTD neuron, with successfully spiking edge detections at three instances, 20, 60 and 120\,ns (highlighted in grey colour). Figure \ref{fig:ed:example} shows that at 20\,ns, there is a fast rising edge, following a sharp increase in the raw signal value. As a result, the difference between the raw and delayed signals (see Figure \ref{fig:ed:example}(d)) modulating the RTD also experience a large increase. In this situation, the amplitude of the optical input pulses go beyond the firing threshold of the RTD, and the system begins firing spikes. When the delayed signal increases 10\,ns later, the difference between the time-series is once again reduced, resulting in a cease fire of spikes. Here, the jump in the RTD bias voltage increases the spike firing threshold of the system, preventing the same optical input pulses from eliciting spikes. A similar result is observed at time instants 60\,ns, and 120\,ns, for different levels of optical and electrical input where the delayed difference is still sufficient to reach the spike firing regime and detect the fast rising edge features. Where the step in raw signal is small (at 100\,ns), the system is unable to surpass the RTD excitation threshold despite operating near the NDR with a low bias voltage, due to weak optical input pulses. These results illustrate the system’s capability to threshold counteracting, simultaneous, inputs and create an event-based rising edge detection system. This RTD-based system permits the tuning of its spiking threshold at will, using either the strength of optical inputs, the voltage modulation point and range, or both, to realise unsupervised time-series edge detection. 

\begin{figure}
    \centering
    \includegraphics[width=0.7\linewidth]{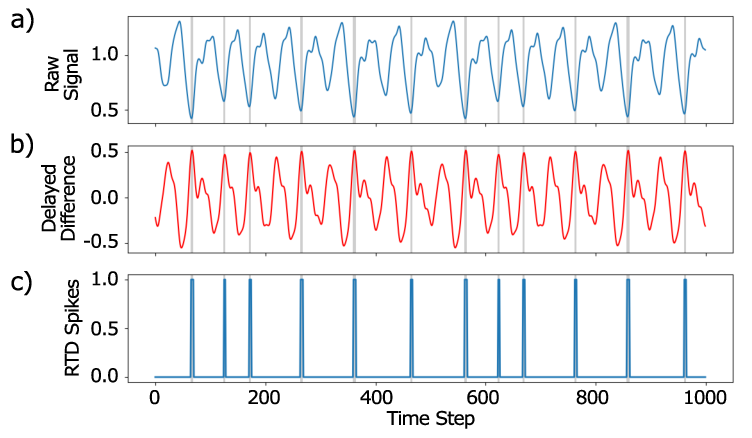}
    \caption{Rising edge detection of the Mackey-Glass time-series performed by a simulated RTD. a) The Mackey-Glass chaotic time-series signal. b) The calculated delayed difference between the raw MG signal and a copy delayed by 10 time steps. c) Time steps for which the RTD emitted a spike ('0' represents no spike, '1' represents a spike), showing that the dual modulation method can reliably detect rising edges. Highlighted in grey (a-c) are regions where edge detections are expected after exceeding a delayed difference of 0.45.}
    \label{fig:ed:sim-results}
\end{figure}

To further test the RTD neuron at this task, we applied it to the detection of rising edges in a complex chaotic time-series, namely the Mackey-Glass (MG) time-series \cite{MackeyGlass}. The latter follows a quasi-periodic pattern and has many regions with differing gradients, hence defining a much more complex rising edge detection task. \textbf{Figure \ref{fig:ed:sim-results}} shows numerical results of a multi-modal RTD spiking neuron tasked with the Mackey-Glass time-series rising edge detection task. Figure \ref{fig:ed:sim-results}(a) shows the Mackey-Glass time-series signal analysed in the study. Figure \ref{fig:ed:sim-results}(b) visualises the delay difference between the raw MG signal, and a copy delayed by 10 time steps. The regions shadowed in grey indicate the rising edges that are targeted by the RTD spiking neuron.

For this case of analysis, the values of the generated MG time-series (featuring an MG tau of $17$, integrated by RK4 with a step size of $0.01\,s$ and sampled every $1\,s$) were first normalised to the range 0 to 1. The raw MG optical input was then created with pulses up to an amplitude of 600 arbitrary units, widths of 20\,ps, and separations of 500\,ps. The delayed copy of the MG time-series was also electrically injected into the RTD neuron. A constant voltage of 0.75\,V was modulated with up to 0.2\,V for a maximum voltage of 1.05\,V. Figure \ref{fig:ed:sim-results}(c) plots the time steps in which the RTD neuron produced spike events (‘1’ spiking, ‘0’ no spiking). These are overlaid by grey shaded regions, showing where edge detections are expected. A threshold of 0.45 delayed difference was found to detect the sharpest rising edges in the MG times-series (see Figure \ref{fig:ed:sim-results}(b)). This threshold value (0.45) was then implemented empirically by correctly choosing the optical pulse amplitude and voltage modulation levels that together produced the matching spiking output. Additionally, we note that changing the delay (currently set to 10 timesteps) also permits the detection of steeper or more shallow rising edge features in the time-series. This parameter is intrinsic to the signal used and we therefore do not rely on it to program the spiking threshold of the system. Once the RTD spiking threshold was correctly tuned, the event-based rising edge detection system was able to correctly identify all target features in the MG time-series. The results in Figure \ref{fig:ed:sim-results} therefore demonstrate numerically that a single multi-modal RTD spiking neuron can be used to successfully implement a rising edge time-series detection system. 

Following the numerical investigation, we experimentally applied a photo-detecting RTD neuron to this rising edge feature detection task. The experimental setup is discussed in \textbf{Section \ref{sec:exp}}. In our experiments, the RTD was biased with a constant voltage of 0.61\,V in the so-called ‘peak’ region. We must note that for the case of biasing the RTD in its peak region, to optically induce spiking responses in the RTD, the optical input signal is flipped. We used a continuous background optical amplitude, with fast drops in optical power to trigger spike-firing events \cite{Qusay2023}. Similarly, we flipped the bias modulation such that a reduction in the bias voltage (moving further from the NDR region) increased the spiking firing threshold and optical amplitude requirement. We encoded the raw MG time-series values in the optical input and the delayed-copy of the signal in the electronic bias modulation. Similar to the numerical simulation, 2\,ns-long (negative) square pulses were used every 200\,ns to encode the MG time-series value. An average optical input power of 2\,mW was used. The electrical modulation varied the RTD with up to -0.3\,V, creating a bias voltage range of 0.61\,V to 0.31\,V, large enough to inhibit even the largest optical inputs. Each RTD bias was again held for 200\,ns. 

\begin{figure}
    \centering
    \includegraphics[width=0.7\linewidth]{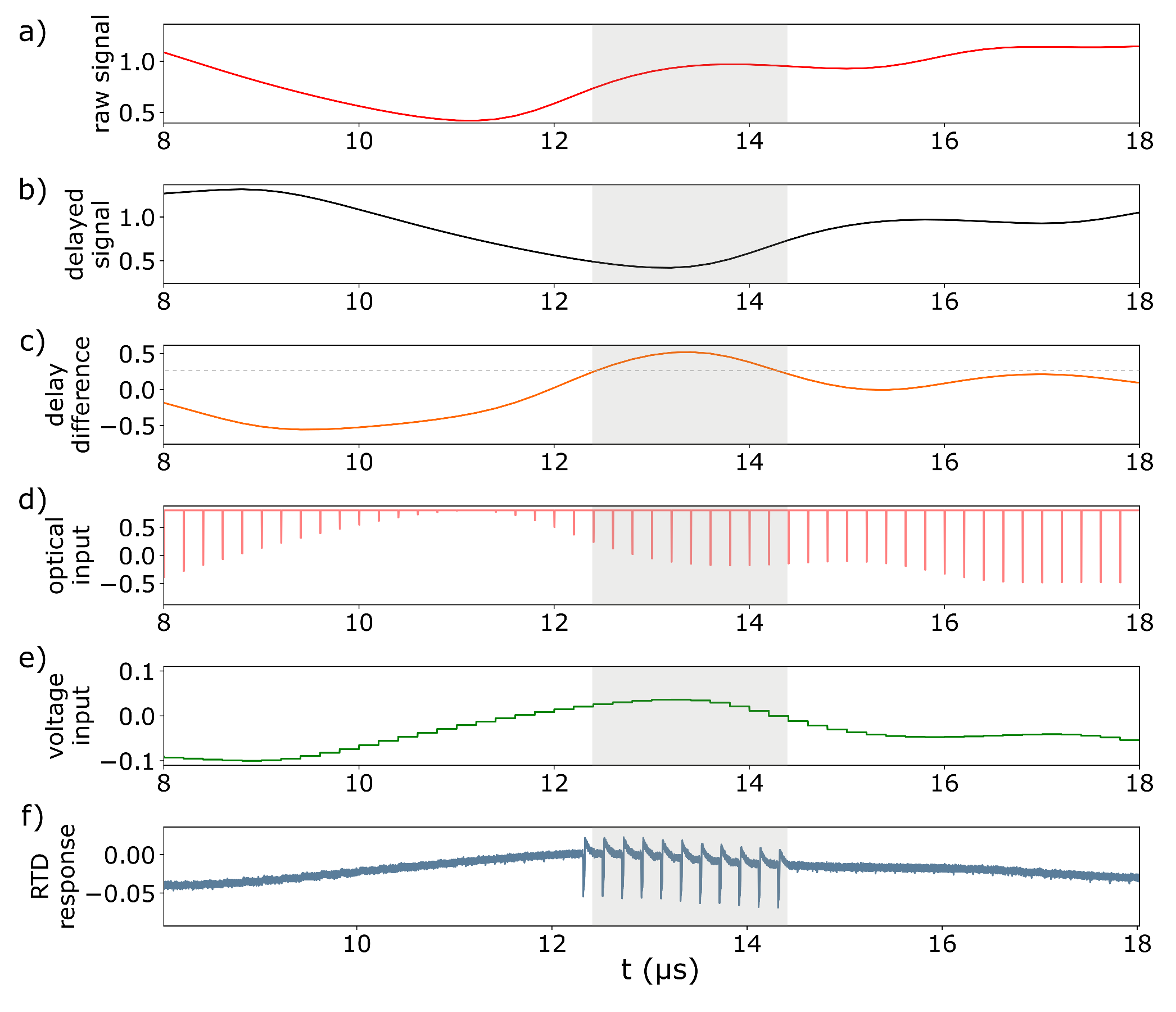}
    \caption{Experimentally measured RTD time-series showing rising edge-feature detection of Mackey-Glass time-series. a) Segment of raw MG time-series data. b) Delayed-copy of MG time-series (delayed by 10 time steps). c) Visualisation of the delayed difference between raw and delayed signals. d) The pulsed optical input used to encode the raw MG timse-series. Pulse widths of 20\,ps are separated by 500\,ps. e) The modulation voltage used to encode the delayed signal into the bias voltage of the RTD. f) Measurement of the experimental (DC-filtered) voltage across the RTD, showing spikes at points when the time-series experiences a rising edge.}
    \label{fig:ed:exp-spikes}
\end{figure}

\begin{figure}
    \centering
    \includegraphics[width=0.7\linewidth]{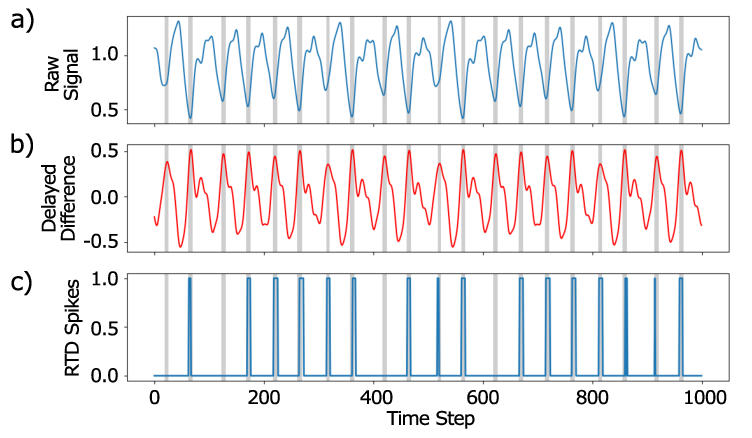}
    \caption{Rising edge detection of the Mackey-Glass time-series performed by an experimental RTD.  a) The Mackey-Glass chaotic time-series signal. b) The calculated delayed difference between the raw MG signal and a copy delayed by 10 time steps. c) Time steps for which the experimental RTD emitted a spike ('0' represents no spike, '1' represents a spike), showing a proof of concept demonstration of dual modulation for rising edge detection. Highlighted in grey (a-c) are regions where detections are expected after exceeding a delayed difference of 0.32.}
    \label{fig:ed:exp-results}
\end{figure}

\textbf{Figure \ref{fig:ed:exp-spikes}} and \textbf{\ref{fig:ed:exp-results}} demonstrate the achievement of the rising edge-feature detection task with the experimental RTD neuron. Figure \ref{fig:ed:exp-spikes}(a) and (b) plot respectively, the raw values of the MG time-series signal and its delayed-copy, using a time-step of 200\,ns. A visualisation of the delay difference in plotted in Figure \ref{fig:ed:exp-spikes}(c). The first signal is optically-encoded into ‘negative’ optical pulses (2\,ns per time-step) as shown in Figure \ref{fig:ed:exp-spikes}(d). Here, higher values in the raw MG time-series signal are encoded with larger amplitude optical pulses. Figure \ref{fig:ed:exp-spikes}(e) plots in turn the delayed values encoded electrically into the bias voltage modulation of the RTD, with higher time-series values encoded as lower voltages. Upon the simultaneous injection of the optical and electrical signals, the light-sensitive RTD produced the temporal spiking output shown in Figure \ref{fig:ed:exp-spikes}(f). The system revealed that excitable spikes were fired in response to the combinations of optical pulse powers and bias voltage inputs that represent a significant delayed difference in MG time-series signals. These points, that identify the rising edge-features in the time-series, occur when the amplitude of the optical input is large and the bias maintains operation near the peak point (closer to 0\,V modulation).

Overall, the experimental results of the task are summarised in Figure \ref{fig:ed:exp-results}. Figure \ref{fig:ed:exp-results}(a) shows the tested Mackey-Glass chaotic time-series, whilst Figure \ref{fig:ed:exp-results}(b) plots the delayed difference between the raw and 10-time step delayed signals. Here, the grey shaded regions reveal time steps where rising edges (where the delayed difference exceeds 0.32), occur in the time-series. The time-series in Figure \ref{fig:ed:exp-results}(c) reveals the time steps for which the RTD fired an excitable spike (‘1’ indicating spiking, '0' indicating non-spiking). Figure \ref{fig:ed:exp-results} reveals that the proof-of-concept experimental demonstration of the system was successful, having detected the majority of the rising edges in the time-series (16 out of 20) in total, without false positive detections. Further, due to the chaotic, slow-varying, nature of the time-series, we would expect to see improvements in performance when applied to real-world time-series that experience larger and more dramatic amplitude variations in short-temporal windows. We believe that this edge-feature detection demonstration indicates the potential of this technique for ultrafast, spike-based alarm-triggering and change-detection systems with RTD neurons.

\subsection{Spatially-extended processing and memory systems with arrays of RTD neurons}

After the analysis of the computational capabilities of single photonic-electronic RTD neurons for real-time spike-based edge detection, we focus now on larger, spatially-extended systems built with arrays of light-sensitive spiking RTD neurons. In the following, we extend our numerical models to account for the simultaneous operation of a growing number of spatially-extended RTD neurons, including architectures with uncoupled and coupled arrays of RTD elements, and describe their potentials for ultrafast neuromorphic photonic information processing and memory systems.

\subsubsection{Photonic Spiking Neural Network (pSNN) with uncoupled RTD neuron arrays}

We start by investigating arrays of photo-detecting RTDs arranged in a simple, uncoupled architecture, independent from one another. We demonstrate numerically that this simplified approach can be used to build an ultrafast neuromorphic photonic spiking neural network (pSNN), able to perform a complex processing task. Specifically, we demonstrate numerically that a small array of only 20 uncoupled photo-detecting RTD neurons can operate as a Photonic Spiking Extreme Learning Machine (ELM)\cite{HUANG2006489} exhibiting accurate operation in a dataset classification task at ultrafast nanosecond speed rates. ELMs are simple neural network architectures, that share many key aspects to reservoir computers \cite{BorghiMassimo2021Rcbo,Rausell2025}, but differ by only enabling information to flow in a feed-forward manner (without recurrence) \cite{Owen-Newns2022,HUANG2006489,Picco2025Eoop}. As such, ELMs are composed of a single input layer (from which information enters the network), a hidden layer (with random untrained connections between nodes), and an output readout layer. In the ELM paradigm (as in reservoir computing) the input and hidden layer weights are random and fixed (set by the  internal physical details of the reservoir device), only requiring the output layer weights to be trained. This is in stark contrast with common approaches based, for example, upon Deep Neural Networks, that require all connections to be trained, therefore requiring complex and costly learning procedures to calculate a very large number of weights. ELMs therefore dramatically reduce the network training requirements, whilst still allowing excellent performance in complex processing tasks (e.g. dataset classification) of widespread interest \cite{Owen-Newns2022,Biasi2023}. Here, we propose for the first time a high-speed ($>$\,GHz) spatially-extended photonic spiking ELM using RTD neurons, thus paving the way towards novel, ultrafast and low-power photonic SNNs with hardware-friendly implementations and highly-reduced training complexity.

\begin{figure}[h]
    \centering
    \includegraphics[width=0.9\linewidth]{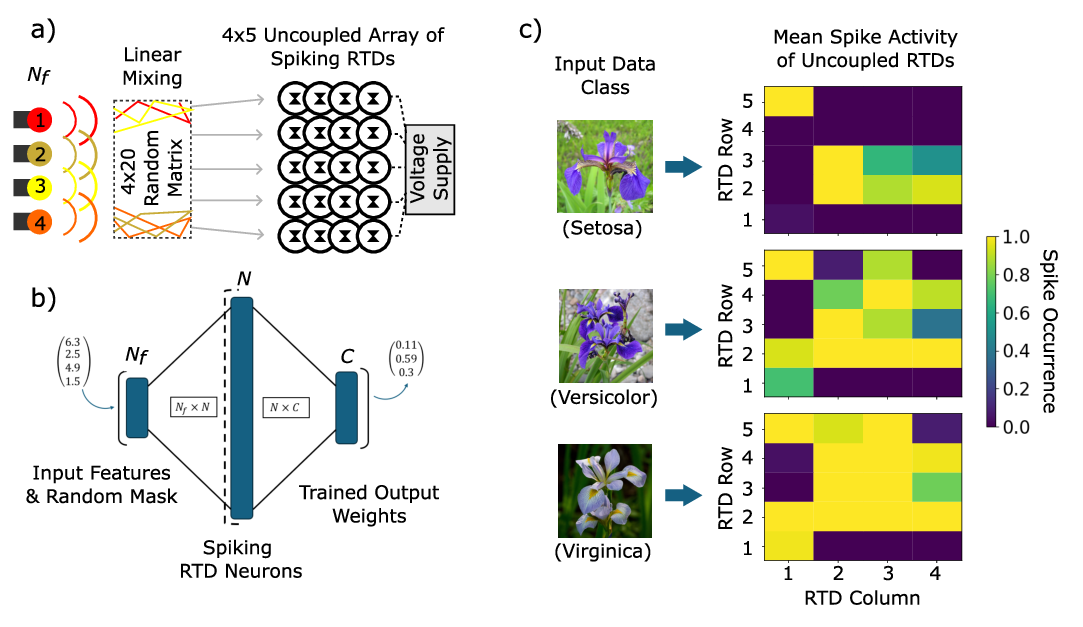}
    \caption{Schematic of the simulated RTD neural network and analysis of RTD-array spikes for the Iris flower dataset classification task. a) Schematic of proposed operation. Iris flower input features (of which there are \textit{$N_f$\,$=$\,$4$}) are randomly scattered creating a linear mixing (weighting) of the inputs. Numerically, linearly mixed inputs are achieved through multiplication with a randomly generated matrix. The mixed inputs arrive optically on each of the (\textit{$N$\,$=$\,$20$}) RTD neurons in the array. b) The architecture of the proposed neural network, including feature mixing/masking (Input layer), RTD neuron array (hidden layer), and the trained readouts (output layer). c) Mean spike activity (occurrence) maps for each dataset class. Spike maps highlight differing responses from spatial RTDs within the array, as well as evolving patterns across dataset classes.}
    \label{fig:ar:setup}
\end{figure}

\textbf{Figure \ref{fig:ar:setup}} shows the operation principle and architecture of the simulated RTD-based photonic spiking ELM proposed in this work. The schematic, shown in Figure \ref{fig:ar:setup}(a), is highly hardware-friendly, with the input data injected optically into the network array of \textit{$N$\,$=$\,$20$} uncoupled spiking RTD neurons. We numerically simulate all 20 RTD neurons with equal properties and parameters (provided in Table \ref{tbl:params}), and bias all devices with the same voltage level of 0.75\,V (in the 'valley' region of their I-V relationship). To demonstrate the neuromorphic processing capability of the proposed pSNN, we investigate the benchmark Iris flower dataset classification task \cite{Fisher:1936}. The task provides 150 datapoints, each containing four input feature values (\textit{$N_f$\,$=$\,$4$}) that identify specimens of three different Iris flower species (Setosa, Versicolor and Virginica). The dataset is equally split with 50 datapoints per class, with 2 of the 3 classes being linearly inseparable from one another. Figure \ref{fig:ar:setup}(b) plots the architecture of the pSNN built with an array of uncoupled RTD neurons. The \textit{$N_f$} data features are encoded using individual amplitude-modulated optical pulses, with a short temporal duration of just 50\,ps. The optical input pulses, encoding the data features, are linearly mixed before entering the 20 uncoupled RTD neurons in the array. Experimentally, this can be achieved for example, by propagating the optical input pulses through a scattering medium prior to their arrival onto the RTD array. We numerically simulate this linear data mixing stage by multiplying the data features (\textit{$N_f$}) with a randomly generated weight matrix ($N_f\,\times\,N$) to produce a weighted input for each of the 20 RTDs. These weighted input values determine the amplitude of the optical pulses used to optically-trigger each spatial photo-detecting RTD in the network. The weighted inputs are injected simultaneously into the photonic spiking ELM at a rate of 1 datapoint per ns, highlighting the high-speed (1\,GHz) neuromorphic processing rate of this proposed system (the rate of data input could be increased further, by several GHz, as allowed by the $\sim300\,ps$ refractory period of the RTD neuron \cite{PhysRevApplied.17.024072}). Spatial spike patterns produced from the 20 RTDs are read-out directly from the simulated electrical output of each RTD neuron in the array. The spike patterns are mapped to binary values ('1' for spiking, '0' for non-spiking) before an output layer weight, calculated following a simple supervised learning approach (linear least squares), is applied to readout the final classification of the datapoint. 

Figure \ref{fig:ar:setup}(c) plots the numerically-calculated mean spiking activity measured at the output of the RTD array, upon the injection of all 50 datapoints from each dataset class. The maps in Figure \ref{fig:ar:setup}(c) show clear differences in the attained spiking patterns from the 20 RTDs for the three different dataset classes, clearly indicating the potential use of spiking activity for classification. In particular, we use use here the photonic spike firing and thresholding capabilities of RTD neurons as an activation function within the feed-forward neural network (photonic ELM). Here, the hidden layer nodes are each represented by a single RTD neuron, with each one of them implementing the Heaviside step function to the amplitude of the input optical pulses:

\begin{align}
    \theta(x-T) = 
    \begin{cases}
    0 & x\leq T \\
      1 & T < x 
    \end{cases}
\end{align}

where, if the input $x$ is greater than the threshold $T$, the RTD will emit a spike. This is then used to construct a spatially multiplexed photonic spiking extreme learning machine (ELM), a feedforward neural network in which only the output layer weights are trained, while the input weights are fixed and selected at random. The small network size, created with a $4\,\times\,5$ array of RTDs, allows for reduced physical complexity and overall power usage, whilst still permitting operation at ultrafast speeds (GHz rates). 

\begin{figure}[h]
    \centering
    \includegraphics[width=0.8\linewidth]{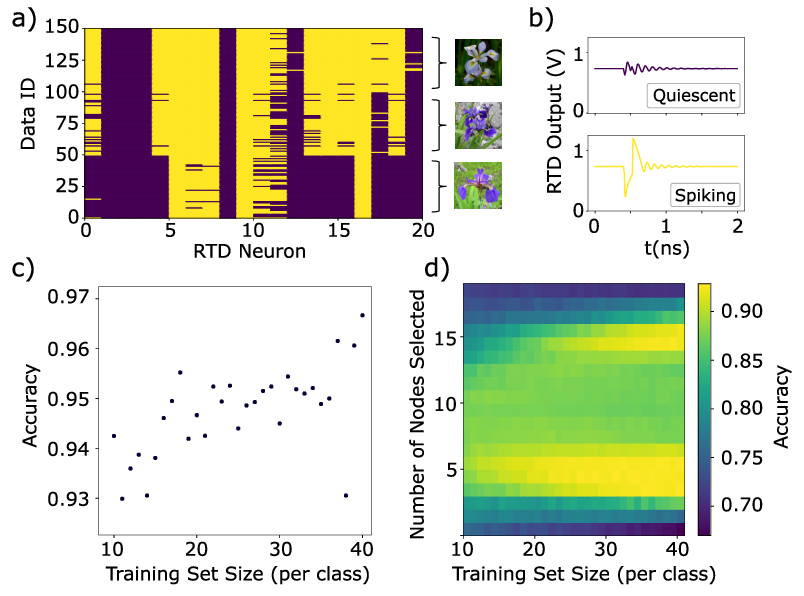}
    \caption{Iris flower dataset classification with a simulated array of RTDs. a) Spiking activity of the photonic spiking ELM generated by each of the 20 uncoupled RTD neurons in the array, for all 150 datapoints of the task (50 per class). Yellow indicates the detection of a spike-firing event, purple denotes no spike firing (quiescence), for the associated node and datapoint. (b) Example time-series of the simulated RTD neurons with spiking and non-spiking responses to datapoints. (c) Accuracy of the RTD ELM when trained with the with least squares method. Maximum accuracy reached was 96.5\%, for 39 points per class in the train set. d) Accuracy when trained with the node significance selection algorithm. Maximum accuracy reached was 93\%, for 35 points per class for training and selecting 6 nodes only.}
    \label{fig:ar:sp}
\end{figure}

The numerically-calculated spiking output of the system, when tasked with the Iris flower dataset, is shown in more detail in \textbf{Figure \ref{fig:ar:sp}}(a). Here, binary responses to the datapoints of each class are shown, revealing that different classes produces slightly different spiking patterns. These plots are produced by representing the spatial ELM as a $1\,\times\,20$ dimensional vector in which each element denotes whether the corresponding RTD neuron in the array spiked or not ('1' for spiking, ‘0’ for non-spiking). Examples of the simulated RTD outputs for the cases of spiking and non-spiking (quiescent) responses are shown in Figure \ref{fig:ar:sp}(b).

After gathering the binary output of the array, learning procedures were applied to train the output layer weights. The first training method used was applying a linear least squares algorithm to the $1\,\times\,20$ vector to calculate an output weight matrix. The results achieved with the least squares training are plotted in Figure \ref{fig:ar:sp}(c), showing the testing accuracy against the number of datapoints of each class used for the training dataset (with the remaining datapoints used for testing). The accuracy achieved with this approach, for the proposed photonic spiking ELM, was consistently above 93\%, even for the cases of using small training set dimensions. Performance reached a peak of 96.5\%, when trained with 39 points of each class, successfully demonstrating the computational potential of the proposed RTD-based photonic spiking ELM.

Furthermore, due to the discrete nature of the spiking responses from the photodetecting RTDs (either spiking or non-spiking), we developed a ‘node significance’ algorithm for the training of binary (0 or 1) weights, recently introduced in \cite{icomputing.0031}. This ‘node significance’ algorithm identifies nodes (spatial RTDs in the array) that contain the most information relevant to the classification task, and sets their weights to ‘1’ for the class that they indicate the strongest. For instance, a node that spikes for all three classes offers no information that can be used to classify an unknown datapoint, whereas a node that spikes persistently only for one class but not the others is very strongly indicative of the points pertaining to that class, and therefore holds high significance. This method introduces a new hyperparameter for the training method, the number of network nodes to select for the training phase, which can have a significant effect on performance (for full details on the node significance algorithm see\cite{icomputing.0031}). The map in Figure \ref{fig:ar:sp}(d) shows the testing accuracy of the photonic spiking ELM when trained using this novel ‘node significance’ algorithm. Performance is measured against the number of points per class in the training set, and against the number of ‘significant nodes’ selected by the algorithm. The peak performance reached was 93\% classification accuracy, for 35 points per class used for training and 6 nodes selected. This value matched the performance of the least squares training algorithm, however produced a more simple, sparse, binary-weight matrix. The increased sparsity, and binary output weights, have positive connotations for the implementation of these systems in hardware, as reduced training calculations and complexity can improve the energy efficiency and training time of the RTD array-based pSNN. 

\subsubsection{Neuromorphic Photonic Memory built with coupled RTD neurons}

Memory in artificial neural networks is typically created by recurrent connections between neurons, in which the output of a neuron is fed back into itself (or, via a longer path of neurons that lead to the original). Recurrent neural networks have found applications in time-series processing task such as speech recognition and language processing \cite{HochreiterSepp1997LSM}, however, they suffer from so called vanishing gradients that lead to slow training. The temporal nature of pSNNs on the other hand, leads to recurrent connections being handled naturally. A regenerative spiking memory system, composed of a single RTD neuron with its optical output connected via delay line to its own optical input, was previously reported \cite{PhysRevApplied.22.024050}, constructing the simplest possible recurrent spiking neural network. In this way, spikes emitted by the RTD-laser neuron were able to repeatedly trigger new spikes after the delay has passed \cite{PhysRevApplied.22.024050,SciRep.6.19510}. This represents a memory system as the state (i.e. the existence and timing of the spike) retains information about the previous inputs to the photonic RTD neuron system.

\begin{figure}[h]
    \centering
    \includegraphics[width=0.95\linewidth]{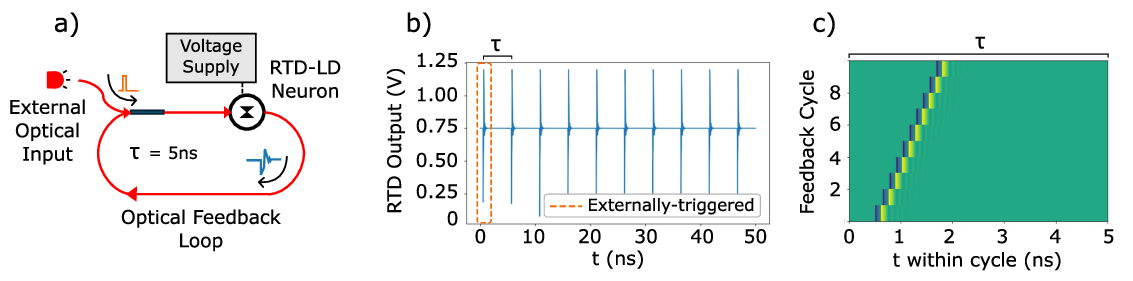}
    \caption{Autaptic spiking memory system with a simulated RTD-laser neuron. a) Schematic of the simulated system. An external optical output is used to initially trigger a single spiking response. The RTD-laser neuron output is fed back to itself via a 5\,ns-long optical delay line. b) Simulated temporal output of the RTD-laser neuron for 10 memory cycles, successfully regenerating the spike each time. c) Map of the temporal response of the RTD-laser neuron for each cycle, showing the an additional spike activation delay of 131\,ps when the spike is regenerated.}
    \label{fig:sm:spikes}
\end{figure}

We start our investigation into spiking photonic memory with RTD-laser neurons by reproducing numerically the operation of this autaptic system. To achieve this we simulate a single RTD-laser neuron connected to itself via a temporal optical feedback loop. \textbf{Figure \ref{fig:sm:spikes}}(a) demonstrates the operation of this autaptic photonic spiking memory, with a delay ($\tau$) of 5 ns, repeatedly storing a single spike event. The RTD-laser neuron is initially triggered by an external optical pulse (dotted box in Figure \ref{fig:sm:spikes}(b)). The RTD-laser neuron fires this first spike in response to a positive optical pulse of 10\,ps duration. After this initial input, the system is only influenced by the delayed optical feedback, where a spike is regenerated by the previous cycle every 5\,ns. Figure \ref{fig:sm:spikes}(c) maps the temporal RTD-laser neuron output for each memory cycle. This map reveals that a slight delay of 131\,ps occurs in each cycle due to the time taken for the RTD-laser neuron to fire a spike (noise in the feedback can also introduce drift in the spike time). We can therefore successfully store spikes within an autaptic spiking system created with an optical feedback loop and an RTD-laser neuron. For analysis of spiking dynamics in similar photonic systems with delayed feedback, see \cite{SciRep.6.19510,TerrienSoizic2020Eopt,TerrienSoizic2021Psbi}.

The limitation of the above single device memory, is that once a spike is introduced to the system, so long as there is no other perturbation, it will continue to be stored indefinitely. For information computing purposes it is preferable to have what is called ”fading memory” \cite{BoydS.1985Fmat,SciRep.2.514,jaeger2002short} - where the effect of any input perturbation eventually fades away to zero, and the system has a single steady state. Here, we introduce and demonstrate numerically a system formed by a network of optically coupled RTD-laser neurons that act similar to a regenerative memory cell, but with finite memory depth (number of cycles information remains in the system), and fully tunable fading memory characteristics.

\begin{figure}[h]
    \centering
    \includegraphics[width=0.9\linewidth]{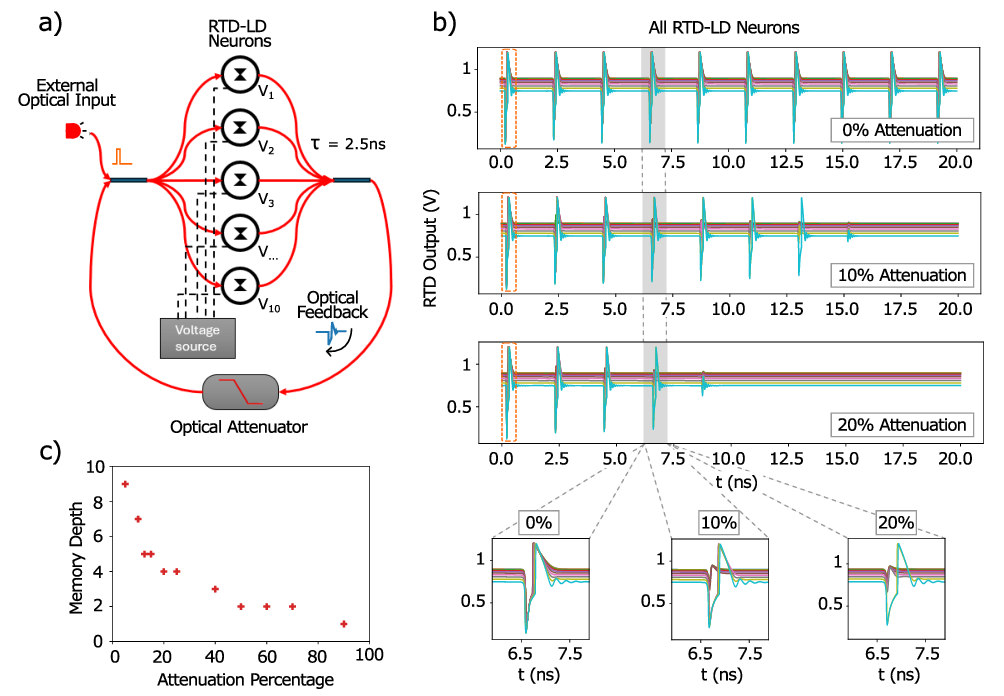}
    \caption{Neuromorphic fading memory cell constructed of 10 RTD-laser neurons. (a) Schematic of the system with RTD-laser neurons connected all-to-all via an attenuator. Each of the RTD-laser neurons are biased at different voltages $V_{0n}=0.9-0.185n^2$, and experience a feedback loop of 2.5\,ns. (b) Simulated RTD-laser neuron voltage traces (10 devices overlaid) for 10 memory cycles. External optical input pulse is used to trigger the first response in each plot (dotted box). Top, middle and bottom plots (and respective insets) show results for increasing optical feedback attenuation (0, 10 and 20\%). (c) Measured memory depth plotted again various feedback attenuation levels showing tunability.}
    \label{fig:mc:demo}
\end{figure}

\textbf{Figure \ref{fig:mc:demo}}(a) shows a schematic of the proposed system used to implement this adjustable neuromorphic photonic fading memory cell. Ten RTD-laser neurons are connected, all-to-all, through fan-in and fan-out optical connections (via optical fibres for example). An optical attenuator is implemented in the (combined) feedback loop such that the optical outputs are weighted equally and combined in a single optical path for reinjection. Each RTD-laser neuron sees the same delayed optical input and the optical power landing on device equals the mean output of all devices (added to any external inputs). This system of 10 simulated photonic-electronic RTD-laser neurons is referred to as a ’memory cell’, as it is a self-contained structure that can exhibit photonic spiking memory. If all RTD-laser neurons in the memory cell were the same (identical features, given the same bias voltage and laser current), it would function identically to a single RTD-laser in a self-feedback loop, where an excitation of one RTD-laser would be continued in the loop indefinitely. For this reason, we drive each device with a different bias voltage. In combination with the attenuated optical feedback, this grants control over the temporal spiking responses of the RTD-laser neurons in the cell, and importantly the overall spiking memory depth. Due to the bias voltage, each RTD-laser neuron in the memory cell will be positioned further/closer from the valley region, creating various spike activation threshold requirements for each neuron. This will cause the RTD-laser neurons to elicit spikes when subject to higher/lower levels of optical feedback (i.e. more/fewer optical spikes stored in memory). Finally, depending on the strength of the optical attenuation, the overall number of devices successfully regenerating spikes into memory can be tuned, providing neuromorphic photonic spiking fading memory cell system.

Figure \ref{fig:mc:demo}(b) plots the numerically-calculated time-series demonstrating the operation of the 10 coupled RTD-laser neuron fading memory cell. Each device in the network was biased at progressively higher voltage, according to the formula $B_n = 0.9-0.185n^2$, such that the lowest was set at 0.75\,V (near the valley point), and the highest at 0.9\,V (far from the NDR with a high spike activation threshold, see Figure \ref{fig:ed:thr} for the variation of threshold with bias voltage). The parabolic distribution of the biases accounts for the non-linear change in spike threshold with bias. The system was set with a delayed feedback loop of 2.5\,ns, and was subject to injection of a single 50\,ps long optical input pulse at $t$ = 10\,ps. This positive optical pulse triggered the firing and storage of the first optical spike event in the network for every RTD-laser neuron. Figure \ref{fig:mc:demo}(b) shows the responses from each RTD-laser neuron (overlaid) when the optical attenuation in the delayed feedback loop was increased from zero to 10\% and 20\%. Figure \ref{fig:mc:demo}(b, top) reveals that all RTD-laser neurons were triggered by the external optical input, and as a result of the high feedback strength, regenerated spikes every 2.5\,ns. This shows that a permanent memory was achieved with a spike always regenerating each cycle, as in the single autaptic neuron case investigated before. For increased attenuation (Figure \ref{fig:mc:demo}(b, middle) and \ref{fig:mc:demo}(b, bottom)), the amplitude of the recurring optical spikes was reduced, therefore a different system memory depth becomes apparent. For the case of 10\% attenuation in Figure \ref{fig:mc:demo}(b, middle), the RTD-laser neurons with the highest bias voltage did not receive enough optical power from the feedback loop to trigger a spike, even when all 10 neurons contributed to the initial input. The number of RTD-laser neurons that fired therefore decreased with each delayed feedback cycle, enabling the achievement of a fading optical memory. Figure \ref{fig:mc:demo}(b, bottom) shows in turn that with even higher attenuation (20\%), the number of firing RTD-laser neurons decreased faster; hence reducing the total memory depth. In the fading memory cell feedback attenuation thus controls the ’depth’ of the fading memory (how long a spike remains in the memory cell). For the case of the 10 device memory cell (under the investigated bias voltage values), further increasing the attenuation decreased the number of feedback cycles until spikes would no longer regenerate. Figure \ref{fig:mc:demo}(c) plots the maximum number of cycles a spike was stored in memory against the optical attenuation strength. With zero attenuation (or any amplification of feedback), the memory depth is infinite, as the spike does not disappear at all. For attenuation over 90\%, the spike is not regenerated at all, even in the first feedback cycle. The relationship between memory depth and attenuation for the studied system was non-linear. Overall, this demonstrates well the ability to tune the amount of time a spike is held in the system, just by adjusting the attenuation, without any change to the physical characteristics of the system (number of RTD-laser neurons, individual connections etc.).

\section{Conclusion}

This work introduces several novel uses of photonic-electronic spiking resonant tunnelling diode neurons for the fast and efficient neuromorphic processing of data, from applications of single device systems, to the implementation of several optically-coupled devices in larger, more complex network architectures. The investigated RTD neurons have shown that they can be made to spike using short pulses of optical power, and importantly, that the required amplitude of these optical pulses is interlinked with the bias voltage applied to the RTD. These parameters give rise to a spike activation threshold that enabled it to be applied to different neuromorphic tasks. The first task demonstrated was rising edge-feature detection in time-series data. This was implemented using the dual modulation of the voltage bias and optical input incident on the RTD, to controllable trigger neuron-like spikes. With each modulation counteracting one-another, we demonstrated the capability to calculate the difference between a raw time-series signal and its delayed copy, such that the RTD emitted spikes when the difference was large (implying a short-term increase in the series data). We demonstrated this task both numerically and experimentally by successfully detecting rising edge-features in the Mackey-Glass chaotic time-series.

In a second neuromorphic photonic processing demonstration, an array of 20 uncoupled RTD neurons were simulated to implement nodes in a feedforward neural network able to operate at GHz rates. The spatial system, based on an architecture known as an extreme learning machine, employed the spike activation threshold of RTDs as a Heaviside step activation function for network nodes to tackle the benchmark Iris flower dataset classification task. In the network, the initial network layer was simulated via a linear mixing of data features, with the hidden layer implemented by passing the optical input to an array of RTD neurons. Only the output weights of the system were trained, via a linear least square algorithm and a novel 'node significance' algorithm. The simulated RTD array was able to successfully achieve classifications accuracies of 96.5\% and 93\% for respective algorithms, with the number of training nodes in the latter as low as 6 (out of 20), while still providing good performance. 

Finally, neuromorphic spiking memory systems, able to operate with high speed input signals, were created by introducing optical feedback loops to systems of RTD neurons. We reproduced a photonic spiking memory system using a single RTD neuron in an autaptic architecture. This device was able to trigger itself to spike following the reinjection of an optical spike, demonstrating a single RTD memory system, that retained spike information indefinitely. Furthermore, we extended the initial autaptic neuron design to a larger neuromorphic spiking memory system formed by a network of multiple coupled devices. We investigated a memory system formed by a network of 10 RTD-neurons interconnected via a feedback loop with controlled attenuation. We demonstrated that this neuromorphic memory cell could control the number of spike regenerated in each memory cycle via optical feedback attenuation. This allowed for tunable control over the spike memory depth of the system, creating a neuromorphic fading-memory cell with potential for further memory-based processing application.

Overall, we have numerically and experimentally demonstrated the applicability of RTD neurons to several novel neuromorphic processing and memory tasks. We believe that these demonstrations hold the technology of RTDs in high regard as exciting candidates for ultrafast, energy-efficient, spiking neuromorphic photonic-electronic devices and systems. Further, given the technologies compatibility with photonic (at key telecom wavelength bands of 1.3\,$\mu m$ and 1.55\,$\mu m$) and high-speed electronic platforms, RTDs have the potential to be key-enabling devices in all manner of future neuromorphic hardware.


\iftrue

\section{Experimental Methods}
\label{sec:exp}

\textbf{Figure \ref{fig:ed:setup}} shows the experimental setup used to investigate the light-sensitive RTD neuron used in this work. The setup featured a bias controller (Keysight E36312) that was used to set the DC voltage bias applied to the RTD. A 12\,GSa/s arbitrary wave generator (AWG, Keysight M8190A 12\,GSa/s) supplied two RF signals used respectively to electrically-modulate the RTD’s voltage bias, and to optically-modulate the light from a tunable 1550\,nm tunable laser source (Santec WSL110) via a 10\,GHz Mach-Zehnder modulator (MZM). The modulated optical signal was injected into the photo-sensitive window of the RTD via a lens-ended optical fibre. The average optical power injected onto the photo-detecting RTD was set to 2\,mW, and the position of the lens-ended fibre was adjusted to achieve maximum coupling efficiency. A coupling loss of 3\,dB was estimated for this system \cite{Zhang2024}. The electrical response of the RTD ws recorded by a real-time oscilloscope (Rohde and Schwartz RTP, 16\,GHz). The modulations were configured such that during rising edge-feature detection the raw MG time-series signal was encoded in the RTD’s optical input, and the delayed-copy electrically modulated the RTD via a bias-tee. 

\begin{figure}[h]
    \centering
    \includegraphics[width=0.4\linewidth]{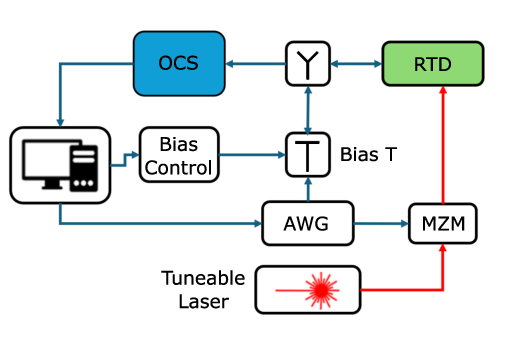}
    \caption{Experimental setup used to probe the photonic-electronic RTD neuron. An AWG modulates the bias of the RTD (through the bias-tee), and the optical power of the signals entering the RTD (via a MZM). The electrical output of the RTD is measured and recorded using a real-time oscilloscope (OSC).}
    \label{fig:ed:setup}
\end{figure}

During the experimental investigation, the RTD was biased at 0.61\,V, at the peak of the I-V curve, shown in Figure \ref{fig:ar:setup} a). A bias point in the peak was chosen due to the increased reliability of spiking response. The electrical modulation was globally scaled, offset and then negated such that all values provided between 0 and -0.3\,V. When added to the bias of 0.61\,V, this created a range of biases in which the RTD would spike given a low amplitude optical input (approx. 0.61\,V), down to biases where not even high amplitude optical pulses would trigger spiking (approx. 0.31\,V). 

\fi

\medskip
\textbf{Acknowledgements} \par 

The authors acknowledge support from the UKRI Turing AI Acceleration Fellowships Programme\\(EP/V025198/1), from the EU Pathfinder Open project ‘SpikePro’ (Grant ID 101129904), and from the Fraunhofer Centre for Applied Photonics, FCAP.

\medskip
\textbf{Conflict of Interest}\\

The Authors declare no conflict of interest.

\medskip
\textbf{Data Availability}

The data used in this work is available from the university of Strathclyde Pure database (doi.org/10.15129/009a7353-6253-4704-979f-669771c96b4d)

%

\medskip

\bibliographystyle{MSP}
\bibliography{bibliography.bib}

\begin{thebibliography}{10}
\providecommand{\url}[1]{\texttt{#1}}
\providecommand{\urlprefix}{URL }

\bibitem{furber2014spinnaker}
S.~B. Furber, F.~Galluppi, S.~Temple, L.~A. Plana,
\newblock \emph{Proceedings of the IEEE} \textbf{2014}, \emph{102}, 5 652.

\bibitem{höppner2022spinnaker2}
S.~Höppner, Y.~Yan, A.~Dixius, S.~Scholze, J.~Partzsch, M.~Stolba, F.~Kelber, B.~Vogginger, F.~Neumärker, G.~Ellguth, S.~Hartmann, S.~Schiefer, T.~Hocker, D.~Walter, G.~Liu, J.~Garside, S.~Furber, C.~Mayr,
\newblock The spinnaker 2 processing element architecture for hybrid digital neuromorphic computing, \textbf{2022},
\newblock \urlprefix\url{https://arxiv.org/abs/2103.08392}.

\bibitem{debole2019truenorth}
M.~V. DeBole, B.~Taba, A.~Amir, F.~Akopyan, A.~Andreopoulos, W.~P. Risk, J.~Kusnitz, C.~O. Otero, T.~K. Nayak, R.~Appuswamy, et~al.,
\newblock \emph{Computer} \textbf{2019}, \emph{52}, 5 20.

\bibitem{Modha2023Northpole}
D.~S. Modha, F.~Akopyan, A.~Andreopoulos, R.~Appuswamy, J.~V. Arthur, A.~S. Cassidy, P.~Datta, M.~V. DeBole, S.~K. Esser, C.~O. Otero, J.~Sawada, B.~Taba, A.~Amir, D.~Bablani, P.~J. Carlson, M.~D. Flickner, R.~Gandhasri, G.~J. Garreau, M.~Ito, J.~L. Klamo, J.~A. Kusnitz, N.~J. McClatchey, J.~L. McKinstry, Y.~Nakamura, T.~K. Nayak, W.~P. Risk, K.~Schleupen, B.~Shaw, J.~Sivagnaname, D.~F. Smith, I.~Terrizzano, T.~Ueda,
\newblock \emph{Science} \textbf{2023}, \emph{382}, 6668 329.

\bibitem{davies2018loihi}
M.~Davies, N.~Srinivasa, T.-H. Lin, G.~Chinya, Y.~Cao, S.~H. Choday, G.~Dimou, P.~Joshi, N.~Imam, S.~Jain, et~al.,
\newblock \emph{Ieee Micro} \textbf{2018}, \emph{38}, 1 82.

\bibitem{Orchard2021Loihi2}
G.~Orchard, E.~P. Frady, D.~B.~D. Rubin, S.~Sanborn, S.~B. Shrestha, F.~T. Sommer, M.~Davies,
\newblock In \emph{2021 IEEE Workshop on Signal Processing Systems (SiPS)}. IEEE,
\newblock ISBN 978-1-6654-0144-9, \textbf{2021} 254--259,
\newblock \urlprefix\url{https://ieeexplore.ieee.org/document/9605018/}.

\bibitem{Brunner2025}
D.~Brunner, B.~J. Shastri, J.~Robertson, X.~Porte, A.~Hurtado,
\newblock Roadmap on neuromorphic photonics : neuromorphic photonics with vertical cavity surface emitting lasers (vcsels), \textbf{2025},
\newblock \urlprefix\url{https://arxiv.org/abs/2501.07917}.

\bibitem{ZhangH2021}
H.~Zhang, M.~Gu, X.~D. Jiang, J.~Thompson, H.~Cai, S.~Paesani, R.~Santagati, A.~Laing, Y.~Zhang, M.~H. Yung, Y.~Z. Shi, F.~K. Muhammad, G.~Q. Lo, X.~S. Luo, B.~Dong, D.~L. Kwong, L.~C. Kwek, A.~Q. Liu,
\newblock \emph{Nature Communications} \textbf{2021}, \emph{12}, 1 457.

\bibitem{Giron2024}
B.~J. Giron~Castro, C.~Peucheret, D.~Zibar, F.~Da~Ros,
\newblock \emph{Optics express} \textbf{2024}, \emph{32}, 2 2039.

\bibitem{Donati2022}
G.~Donati, C.~R. Mirasso, M.~Mancinelli, L.~Pavesi, A.~Argyris,
\newblock \emph{Optics express} \textbf{2022}, \emph{30}, 1 522.

\bibitem{Feldmann2019}
J.~Feldmann, N.~Youngblood, C.~D. Wright, H.~Bhaskaran, W.~H.~P. Pernice,
\newblock \emph{Nature} \textbf{2019}, \emph{569}, 7755 208.

\bibitem{Zheng2023Edoc}
D.~Zheng, S.~Xiang, X.~Guo, Y.~Zhang, B.~Gu, H.~Wang, Z.~Xu, X.~Zhu, Y.~Shi, Y.~Hao,
\newblock \emph{Photonics research (Washington, DC)} \textbf{2023}, \emph{11}, 1 65.

\bibitem{Jacob2025}
B.~Jacob, J.~Silva, J.~M.~L. Figueiredo, J.~B. Nieder, B.~Romeira,
\newblock \emph{Scientific reports} \textbf{2025}, \emph{15}, 1 6805.

\bibitem{Zhang2024}
W.~Zhang, M.~Hejda, Q.~R.~A. Al-Taai, D.~Owen-Newns, B.~Romeira, J.~M.~L. Figueiredo, J.~Robertson, E.~Wasige, A.~Hurtado,
\newblock \emph{Neuromorphic computing and engineering} \textbf{2024}, \emph{4}, 4 44006.

\bibitem{cimbri2022resonant}
D.~Cimbri, J.~Wang, A.~Al-Khalidi, E.~Wasige,
\newblock \emph{IEEE Transactions on Terahertz Science and Technology} \textbf{2022}, \emph{12}, 3 226.

\bibitem{nishida2019terahertz}
Y.~Nishida, N.~Nishigami, S.~Diebold, J.~Kim, M.~Fujita, T.~Nagatsuma,
\newblock \emph{Scientific reports} \textbf{2019}, \emph{9}, 1 18125.

\bibitem{robertson2025leaky}
J.~Robertson, D.~Black, G.~Donati, Q.~R.~A. Al-Taai, E.~Malysheva, B.~Romeira, J.~Figueiredo, V.~D. Calzadilla, E.~Wasige, A.~Hurtado,
\newblock Ultrafast and compact photonic-electronic leaky integrate-and-fire circuits based upon resonant tunnelling diodes,
\newblock \urlprefix\url{https://arxiv.org/abs/2501.17133},
\newblock Accessed: \textbf{2025}.

\bibitem{ZhangW:PRTD}
W.~Zhang, A.~Al-Khalidi, J.~Figueiredo, Q.~R.~A. Al-Taai, E.~Wasige, R.~H. Hadfield,
\newblock \emph{Nanomaterials} \textbf{2021}, \emph{11}, 6 1590.

\bibitem{PhysRevApplied.15.034017}
I.~Ortega-Piwonka, O.~Piro, J.~Figueiredo, B.~Romeira, J.~Javaloyes,
\newblock \emph{Phys. Rev. Appl.} \textbf{2021}, \emph{15} 034017.

\bibitem{Donati2024}
G.~Donati, D.~Owen-Newns, J.~Robertson, E.~Malysheva, A.~Adair, J.~Figueiredo, B.~Romeira, V.~Dolores-Calzadilla, A.~Hurtado,
\newblock \emph{Physical review letters} \textbf{2024}, \emph{133}, 26 267301.

\bibitem{Robertson2024Utip}
J.~Robertson, D.~Black, Q.~R.~A. Al-Taai, G.~Donati, E.~Malysheva, B.~Romeira, J.~Figueiredo, V.~Dolores-Calzadilla, E.~Wasaige, A.~Hurtado,
\newblock In \emph{2024 IEEE Photonics Conference (IPC)}. \textbf{2024} 1--2.

\bibitem{PhysRevApplied.17.024072}
M.~Hejda, J.~A. Alanis, I.~Ortega-Piwonka, J.~a. Louren\ifmmode~\mbox{\c{c}}\else \c{c}\fi{}o, J.~Figueiredo, J.~Javaloyes, B.~Romeira, A.~Hurtado,
\newblock \emph{Phys. Rev. Appl.} \textbf{2022}, \emph{17} 024072.

\bibitem{Qusay2023}
Q.~R.~A. Al-Taai, M.~Hejda, W.~Zhang, B.~Romeira, J.~M.~L. Figueiredo, E.~Wasige, A.~Hurtado,
\newblock \emph{Neuromorphic computing and engineering} \textbf{2023}, \emph{3}, 3 34012.

\bibitem{RTD_IV}
J.~Schulman, H.~De~Los~Santos, D.~Chow,
\newblock \emph{IEEE Electron Device Letters} \textbf{1996}, \emph{17}, 5 220.

\bibitem{10.1063/1.5022958}
J.~Mork, G.~L. Lippi,
\newblock \emph{Applied Physics Letters} \textbf{2018}, \emph{112}, 14 141103.

\bibitem{IserlesA1996Afci}
A.~Iserles,
\newblock \emph{A first course in the numerical analysis of differential equations},
\newblock Cambridge texts in applied mathematics. Cambridge University Press, Cambridge ; New York, \textbf{1996}.

\bibitem{MackeyGlass}
M.~C. Mackey, L.~Glass,
\newblock \emph{Science} \textbf{1977}, \emph{197}, 4300 287.

\bibitem{HUANG2006489}
G.-B. Huang, Q.-Y. Zhu, C.-K. Siew,
\newblock \emph{Neurocomputing} \textbf{2006}, \emph{70}, 1-3 489.

\bibitem{BorghiMassimo2021Rcbo}
M.~Borghi, S.~Biasi, L.~Pavesi,
\newblock \emph{Scientific reports} \textbf{2021}, \emph{11}, 1 15642.

\bibitem{Rausell2025}
J.~R. Rausell‐Campo, A.~Hurtado, D.~Pérez‐López, J.~Capmany~Francoy,
\newblock \emph{Laser \& photonics reviews} \textbf{2025}, \emph{19}, 9.

\bibitem{Owen-Newns2022}
D.~Owen-Newns, J.~Robertson, M.~Hejda, A.~Hurtado,
\newblock \emph{IEEE Journal of Selected Topics in Quantum Electronics} \textbf{2023}, \emph{29}, 2: Optical Computing 1.

\bibitem{Picco2025Eoop}
E.~Picco, L.~Jaurigue, K.~Lüdge, S.~Massar,
\newblock \emph{Communications engineering} \textbf{2025}, \emph{4}, 1 3.

\bibitem{Biasi2023}
S.~Biasi, R.~Franchi, L.~Cerini, L.~Pavesi,
\newblock \emph{APL photonics} \textbf{2023}, \emph{8}, 9 096105.

\bibitem{Fisher:1936}
R.~A. Fisher,
\newblock \emph{Annals of Eugenics} \textbf{1936}, \emph{7}, 2 179.

\bibitem{icomputing.0031}
D.~Owen-Newns, J.~Robertson, M.~Hejda, A.~Hurtado,
\newblock \emph{Intelligent Computing} \textbf{2023}, \emph{2} 0031.

\bibitem{HochreiterSepp1997LSM}
S.~Hochreiter, J.~Schmidhuber,
\newblock \emph{Neural computation} \textbf{1997}, \emph{9}, 8 1735.

\bibitem{PhysRevApplied.22.024050}
J.~M. Martins, S.~V. Gurevich, J.~Javaloyes,
\newblock \emph{Phys. Rev. Appl.} \textbf{2024}, \emph{22} 024050.

\bibitem{SciRep.6.19510}
B.~Romeira, R.~Avó, J.~M.~L. Figueiredo, S.~Barland, J.~Javaloyes,
\newblock \emph{Scientific Reports} \textbf{2016}, \emph{6} 19510.

\bibitem{TerrienSoizic2020Eopt}
S.~Terrien, V.~A. Pammi, N.~G.~R. Broderick, R.~Braive, G.~Beaudoin, I.~Sagnes, B.~Krauskopf, S.~Barbay,
\newblock \emph{Physical review research} \textbf{2020}, \emph{2}, 2 023012.

\bibitem{TerrienSoizic2021Psbi}
S.~Terrien, V.~A. Pammi, B.~Krauskopf, N.~G.~R. Broderick, S.~Barbay,
\newblock \emph{Physical review. E} \textbf{2021}, \emph{103}, 1 012210.

\bibitem{BoydS.1985Fmat}
S.~Boyd, L.~Chua,
\newblock \emph{IEEE transactions on circuits and systems} \textbf{1985}, \emph{32}, 11 1150.

\bibitem{SciRep.2.514}
J.~Dambre, D.~Verstraeten, B.~Schrauwen, S.~Massar,
\newblock \emph{Scientific Reports} \textbf{2012}, \emph{2} 514.

\bibitem{jaeger2002short}
H.~Jaeger,
\newblock \emph{GMD-German National Research Institute for Computer Science (2002)} \textbf{2002}.

\end{thebibliography}


\end{document}